**Experimental and computational dosimetry in an integrated workflow for remote audits in Ir-192 interstitial brachytherapy: Development and pilot implementation**

Eleftherios P Pappas[1], Vasiliki Peppa[1,2], Alexandra Drakopoulou[1], Eleni Velissariou[1], Zoi Thrapsanioti[3], Georgios Kollias[4], Efi Koutsouveli[4], Georgia Lymperopoulou[5], Pantelis Karaiskos[1]

[1]*Medical Physics Laboratory, Medical School, National and Kapodistrian University of Athens, 115 27 Athens, Greece*

[2]*Department of Radiotherapy, Alexandra General Hospital of Athens, 115 28 Athens, Greece*

[3]*Ionizing Radiation Unit, Greek Atomic Energy Commission (EEAE), 153 41 Agia Paraskevi, Greece*

[4]*Medical Physics Department, Hygeia Hospital, 151 23 Marousi, Athens, Greece*

[5]*1st Department of Radiology, Medical School, National and Kapodistrian University of Athens, 115 28 Athens, Greece*

**Keywords:** brachytherapy, dosimetry, film, OSLD, Monte Carlo, audit

**Abstract**

**Purpose:** This work presents the development and pilot implementation of a comprehensive remote (postal) dosimetry audit for Ir-192 High Dose Rate interstitial brachytherapy, integrating independent experimental and computational dosimetry procedures into a unified workflow.

**Methods:** A compact, water-equivalent phantom was designed to accommodate two plastic catheters, ten Optically Stimulated Luminescent Dosimeters (OSLDs) and two radiochromic films, allowing for independent point-approximating and 2D dose measurements. In the pilot study, a user-selected treatment plan (36 source dwell positions) was generated using a clinical Treatment Planning System (TPS), after considering the optimal dose range of the dosimeters. By analyzing the dicom-RT files, a computational dosimetry audit test was also performed using Monte Carlo (MC) simulations, enabling independent 3D dose calculations for the same plan and phantom geometry. All dosimetry results were compared to TPS calculations (TG43 and a Model Based Dose Calculation Algorithm, MBDCA) using the 3D Gamma Index (GI) test, dose difference maps, and dose-volume histogram comparisons, wherever applicable. All procedures were designed for minimum clinical workload burden.

**Results:** The pilot study was completed within ten days of phantom delivery to the clinical site. If necessary, measurements were corrected by applying appropriate correction factors determined by conducting side studies. GI passing criteria were adapted to the uncertainty of each dosimetry system. Excellent agreement between MBDCA dose predictions and experimental or MC results was observed. Within the volume of interest, a systematic overestimation by TG43 relative to MC results (median difference: +2.16%) was attributed to missing scatter conditions and phantom material.

**Conclusion:** Despite the labor-intensive workflow for the auditing institution, the developed protocol is suitable for remote Ir-192 audits with acceptable uncertainties. Combining experimental and computational methods strengthens the reliability of audit outcomes. Overall results of this work highlight the advantages of an integrated dosimetry protocol for comprehensive and rigorous auditing programs.

## 1. INTRODUCTION

Brachytherapy is a well-established radiotherapy treatment modality, which relies on a radioactive isotope encapsulated in a small-scale source for radiation dose delivery to the target. In some of the interstitial high dose rate (HDR) brachytherapy applications available, an Ir-192 source is driven through plastic catheters implanted inside the treatment site of the patient.[1] Treatment planning and optimization are carried out by dedicated Treatment Planning Systems (TPSs), incorporating dose calculation engines based on either the American Association of Physicists in Medicine (AAPM) TG43 protocol,[1] or advanced Model

Based Calculation Algorithms (MBDCAs) according to the AAPM TG186 report.[2] Treatment delivery is facilitated by a remote afterloader. Efficiency of the therapy relies on the accuracy and precision of all components involved in the treatment workflow. Rigorous Quality Assurance (QA) programs are necessary to verify that the calculated dose distribution is delivered to the patient within predefined spatial and dosimetric tolerances.[2, 3]

External audit tests and especially remote (postal) dosimetry tests have become immensely popular in advanced external beam radiotherapy modalities,[4–7] with a notably increasing interest in brachytherapy applications.[8–13] The scope of such procedures is to externally validate the dose delivery accuracy without the physical presence of an auditor or the shipment of special, expensive or large equipment. Experimental dosimetry tests have always been the backbone of any QA procedure in radiotherapy while computational dosimetry tests are more suitable for TPS commissioning or for performance evaluation based on test cases, targeted to the dose calculation engine specifically.[14–17]

With respect to the former, Dimitriadis et al. recently published a methodology for auditing purposes in brachytherapy based on a single Radio Photo Luminescence Dosimeter (RPLD) for absolute dose determination combined with a gafchromic film for verification of the source step size.[12] All components were fit in a small-scale phantom made of acrylic. This audit test, developed by the International Atomic Energy Agency (IAEA), mainly focuses on the verification of the reference air kerma strength.[12] Two Optically Stimulated Luminescence Dosimeters (OSLDs) were used in a polystyrene ($\rho$=1.04 g/cm$^3$) phantom in the audit test developed by the Radiological Physics Center (RPC).[13] Oliver-Canamas et al. also presented a phantom similar in concept but added radiochromic films.[9, 18] A dose-response characterization for all dosimeters and materials involved was performed specifically for the Ir-192 beam quality. Relevant correction factors can also be determined with the aid of Monte Carlo (MC) simulations.[19] Such audit procedures are rigorous but involve a predefined simple treatment plan, and thus cannot be considered as end-to-end verification tests, in lack of treatment planning and optimization steps or clinically relevant dose gradients.

Based on simulations, a computational dosimetry audit test focuses on the accuracy of the TPS and the corresponding dose calculation engine, while mitigating increased experimental uncertainties, unexpected errors during implementation or sub-optimal performance of the equipment used for detector readout. Equally important, a computational dosimetry test allows for an inherently 3D treatment plan verification procedure[14, 15, 17, 20] which is challenging in absolute dose terms using 3D experimental dosimetry (e.g., gel).[21, 22] Nevertheless, if a computational dosimetry test is solely implemented, potential errors related to the remote afterloader, the file transfer and communication systems, the source transit time, the source air kerma strength or even the catheter reconstruction and digitization processes may go undetected.

Acknowledging the pros and cons of experimental and computational dosimetry, the main scope of this study is to develop and implement a comprehensive methodology for remote dosimetry audits, leveraging the advantages of both, in an integrated workflow. Regarding measurements, two independent dosimetry systems are employed allowing for point-approximating measurements and 2D dose maps. TPS calculations are also compared in 3D with MC results in a computational dosimetry study for the same case and geometry. While currently available audit tests mainly focus on point dose-to-water validation for a simple predefined plan,[12, 13, 19] this study presents an end-to-end dosimetry test, involving all steps of the treatment chain. An important feature of the methodology developed in the present study is that the experimental dosimetry test is independent of the computational one, while the workflow is integrated into a single procedure, referring to the same plan, geometry, and setup. This approach increases confidence in the results. Wherever necessary, appropriate correction factors are determined in specially designed side studies. Feasibility is demonstrated in a pilot study by implementing the integrated dosimetry protocol in a real-world clinical setting in order to identify potential limitations and quantify the relevant uncertainties during implementation.

## 2. MATERIALS AND METHODS

### 2.1 Overview of the comprehensive remote dosimetry audit test

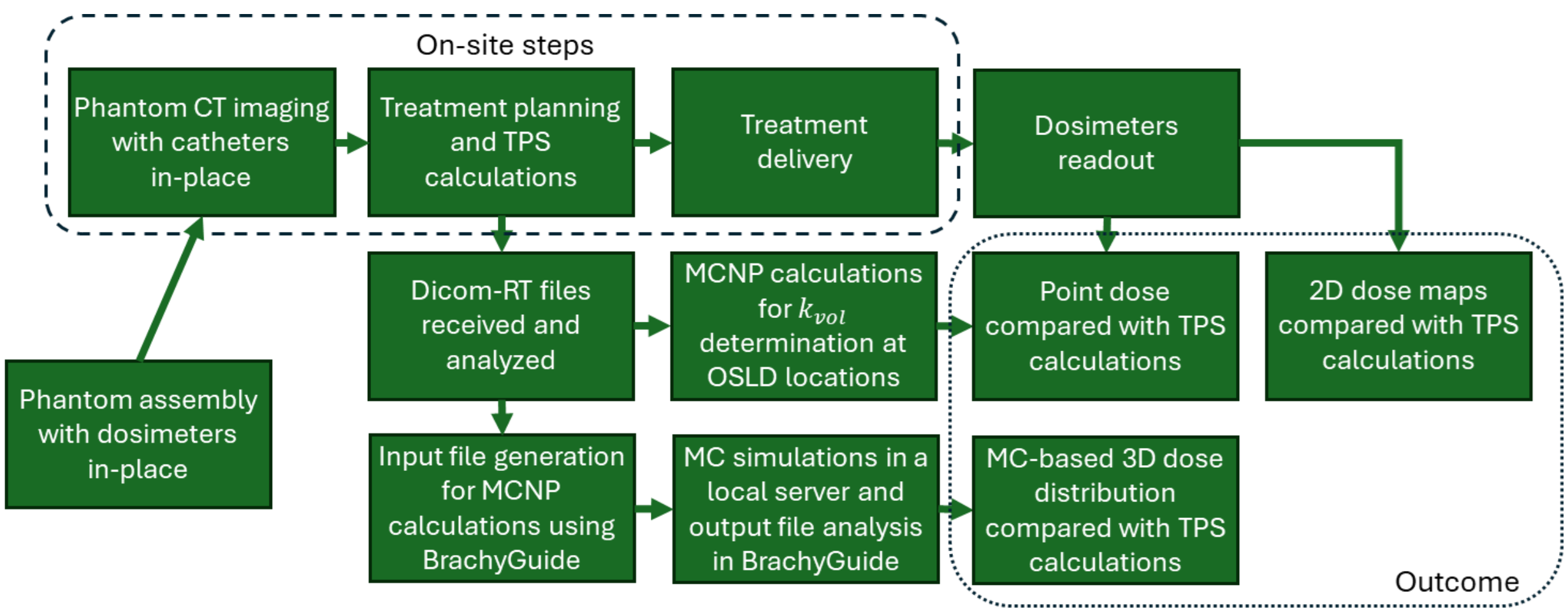

*Figure 1. Overview of the integrated workflow for the developed hybrid dosimetry remote audit test. Abbreviations: CT: Computed Tomography; TPS: Treatment Planning System, OSLD: Optically Stimulated Luminescence Dosimeter; MC: Monte Carlo.*

The workflow of the developed comprehensive audit test is outlined in Figure 1. The key feature is to implement two independent remote dosimetry tests, an experimental and a computational one, in a single integrated protocol, for the same user-selected treatment plan. Although it involves several steps, only phantom imaging, treatment planning and delivery are carried out on-site (Figure 1). All other steps are performed outside the clinical environment and thus without disruption of the patient flow. Moreover, the necessary equipment (phantom and passive detectors) is designed to be compact in size and shape, enabling postal dosimetry capabilities. Details specific to each step, including equipment and methodology, are given in the next sections, implementing the workflow in Figure 1 in a pilot study.

## 2.2 Phantom design and construction

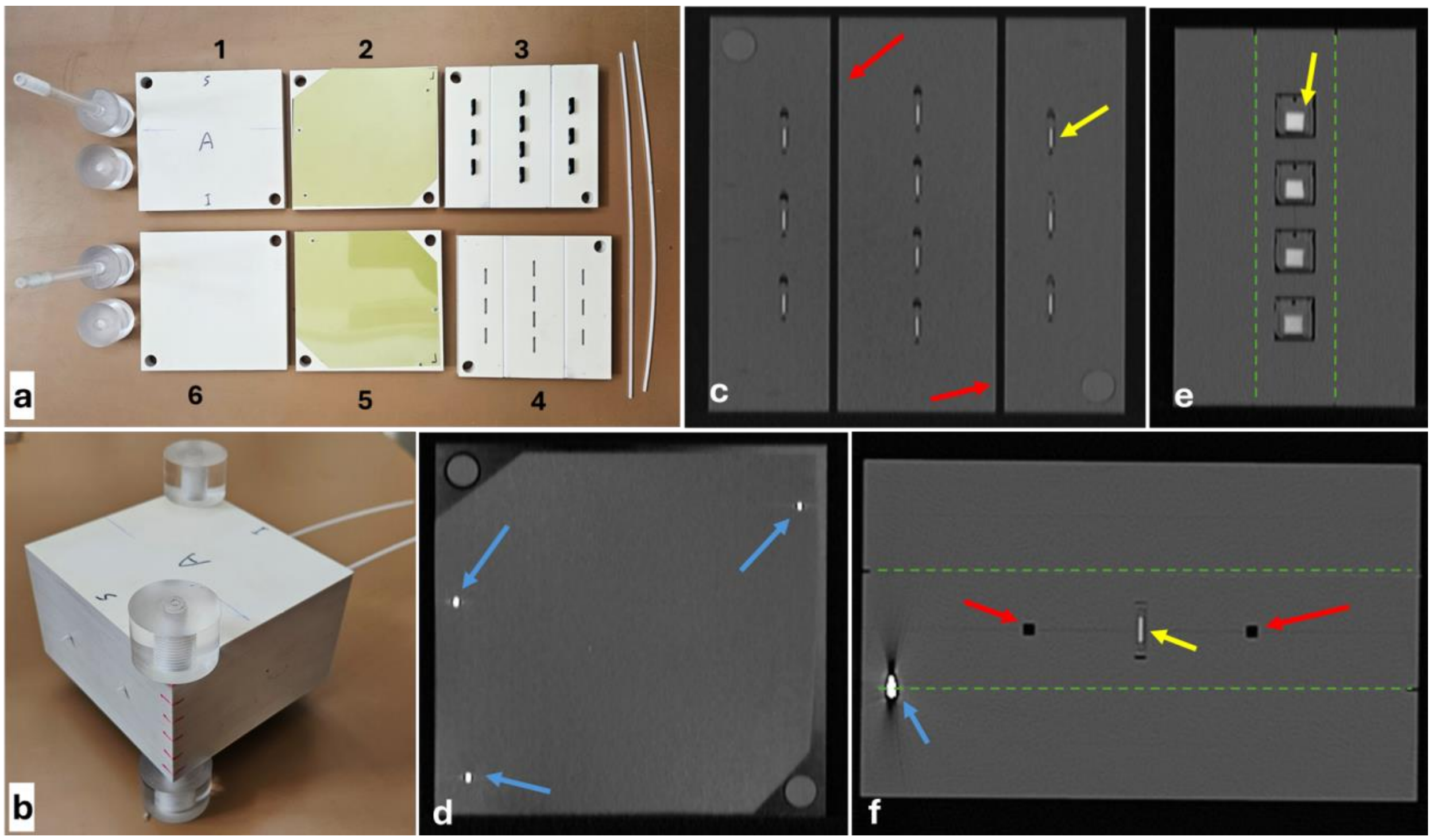

*Figure 2. (a) All six slabs that make up the phantom assembly along with the two plastic catheters and the fixing rods. Ten OSLDs and two films are in-place for demonstration purposes. (b) The phantom assembly incorporating all detectors and two catheters. (c,d) Coronal, (e) sagittal, and (f) axial slices of the reconstructed CT image stack. The plane shown in panel (c) is the interface between slabs 3 and 4 on which both catheters and all ten OSLD centroids lie. The plane shown in panel (d) coincides with one of the two film planes (interface between slabs 2 and 3); the three fiducials for spatial registration of the film dose maps to the CT coordinate system are shown. Legend: yellow, red, and blue arrows point to OSLDs active volumes, catheter channels and metal pins (fiducials), respectively. The green dashed lines represent film planes (interfaces between slabs 2-3 and 4-5). Images are not drawn to scale.*

A dosimetry phantom for interstitial brachytherapy applications was designed and constructed (Figure 2), according to the following characteristics and specifications:

- It is made of a water-equivalent material for the beam quality of Ir-192. In particular, the phantom was constructed by machining a total of six 1-cm thick RW3 slabs (PTW, Germany),[23] as shown in Figure 2a. RW3 can be considered water-equivalent as deviations from dose in a water phantom do not exceed 0.8% up to 10 cm from the Ir-192 source.[24]
- External dimensions (9.5×10×6 $cm^3$, Figure 2b and Table 1a) are within the range in which RW3 can be considered water equivalent.[24] The phantom's size is relevant to breast or tongue cancer cases, but its shape is not anthropomorphic (Figure 2b). Its compact design is suitable for postal dosimetry audit tests. Moreover, this size does not result in an unacceptably large dose calculation grid to achieve adequate accuracy in MC dose calculations. The selected phantom size is also suitable for dose measurements and calculations at clinically relevant distances from the source dwell positions (Table 1a).
- The phantom offers multi detector dosimetry capabilities. In particular, up to ten OSLDs can be used for high-sensitivity point-approximating dose determination and up to two films for high spatial resolution dose distribution measurements (Figure 2). The locations of the detectors were selected in order to match the optimal sensitivity and dose range of each dosimeter type. Employing two fundamentally different dosimetry systems allows for independent dose measurements, enhancing the credibility of the measurements.
- It can incorporate up to two parallel plastic catheters, 2 mm in diameter, by drilling two channels on the surfaces of two paired slabs (Figures 2a-c,f). Consequently, the midpoint of each channel is always found on the interface of slabs 3 and 4 as labeled in Figure 2a.
- All OSLDs are located on the plane defined by the two catheters (slabs 3 and 4 interface, Figure 2a). The in-plane OSLD spatial distribution is symmetrical with respect to the two catheters and the geometrical centroid of the slab's surface (Figure 2c).
- Single film pieces are placed in-between slabs 2-3 and 3-4 (Figure 2a). The normal distance between the catheters' plane and each film plane is 1 cm (Table 1a). More specifically, the film planes lie posteriorly and inferiorly to the catheters' plane, covering most of the area defined by the slabs. Positioning the two films symmetrically with respect to the catheters (Figures 2e-f) allows for two independent 2D dose measurements to be performed concurrently.

- The phantom is designed to accommodate all critical components (dosimeters and catheters) at fixed and well-defined positions in the planning CT scan image series (Figures 2c-f).
- The measured 2D dose maps can be spatially registered to the planning CT coordinate system, based on three fiducial markers (metal pins) on each film (Figures 2d,f). Regarding the OSLDs, a direct calculation of the geometric centroid of the active volume is performed in the CT image space, exploiting the increased Hounsfield Units (HUs) of the active material (Figures 2c,e,f).
- The phantom incorporating all detectors and both catheters is spatially rigid by mounting two rods that lance through all RW3 planes via a snug fit, while two acrylic nuts ensure tight fixing of all components (Figures 2a-b).

Machining of the phantom was performed by a fully robotized Computerized Numerical Control (CNC) router after developing all designs in CAD format. According to its technical specifications, this system offers spatial accuracy of <0.05 mm in cutting and incising in all three dimensions. Moreover, flow of coolant liquid during machining ensures spatial fidelity to the original design by avoiding melting-related inaccuracies. The spatial accuracy of all components of the constructed phantom was verified by using a vernier.

*Table 1. (a) Key specifications of the developed brachytherapy dosimetry phantom. (b) Treatment planning and delivery details of the pilot study.*

| | | |
|---|---|---|
| (a) | Phantom external dimensions | 9.5×10×6 $cm^3$ |
| | Material | RW3 (ρ=1.04 $g/cm^3$) |
| | Cather-to-catheter distance | 4 cm |
| | Catheter-to-OSLD distance | 1.2 or 2 cm |
| | OSLD-to-OSLD min distance | 1.4 or 1.7 cm |
| | Film-to-catheter distance | 1 cm |
| | Film-to-film distance | 2 cm |
| (b) | Number of catheters | 2 |
| | Number of active dwell positions | 36 |
| | Range of dwell times | 5.1 – 15.8 sec |
| | Source model | FlexHDR20ch (Elekta Brachy, The Netherlands) |
| | Afterloader | Flexitron HDR 192-Ir (Elekta Brachy, The Netherlands) |
| | Treatment Panning System | Oncentra Brachy v.4.6 (Elekta Brachy, The Netherlands) |
| | Overall treatment duration | 414.3 sec |

### 2.3 Pilot study: imaging and treatment planning

To conduct a pilot study, the phantom was assembled with all detectors and catheters in-place (Figure 2b). The external surface was marked to define nominal orientation for imaging and treatment planning purposes. More specifically, the catheters were parallel to the superior – inferior direction, while the film planes lied anteriorly and posteriorly with respect to the catheters, i.e., film measurements corresponded to a nominal coronal plane (e.g., Figure 2d).

Meticulous phantom positioning and alignment for CT imaging and treatment planning was not necessary, although an oblique scan would result in substantial partial volume effects and thus was avoided. Images were acquired by a SOMATOM Confidence scanner (Siemens Healthineers, Germany) at 120 kV. Reconstructed pixel size was $0.27\times0.27$ mm$^2$ with a slice thickness of 0.6 mm. In addition, CT scanning was repeated after placing wires (x-ray markers) in the catheters to induce high contrast facilitating accurate catheter tip identification and catheter reconstruction. The latter step was performed in the TPS following the standard clinical practice, in case catheter tips are in-air.

Acquired images were imported to the Oncentra Brachy v.4.6 TPS (Elekta Brachy, The Netherlands) for treatment planning. Catheter reconstruction was carried out following the standard clinical workflow. Three hypothetical structures labelled "high dose region", "low dose region 1", and "low dose region 2" in the area in-between the two catheters (red, blue, and turquoise contours, respectively, Figure 3) were considered to assist treatment planning and optimization. Source dwell positions were enabled with a step size of 2 mm (Table 1b) for a treatment length of 3.4 cm (red marks, Figure 3). Given the limited size of the phantom and the limited number of catheters, selecting a step size of 2 mm allowed for a larger number of source dwell positions and, thus, achieving increased dose modulation, if required. Prescription dose was 3 Gy at points on the mid-axis of the two catheters (i.e., located 2 cm from each source dwell position) while the distribution was optimized using distance optimization to avoid extreme dose gradients within the OSLDs' active volumes. The prescription dose was selected so that all detectors (films and OSLDs) receive doses within their optimal dose range, matching their dose-response characteristics and calibration range, while ensuring that both low (e.g., 50%) and high (e.g., 160%) dose areas are measured.

The TG43 based dose calculation algorithm was employed to determine the TPS-calculated dose-to-water in water distribution with a dose grid resolution of $0.5\times0.5\times0.5$ mm$^3$. This standard calculation algorithm does not consider material inhomogeneities while full scatter photon conditions are assumed.[1] For the same plan, dose was also calculated using the Advanced Collapsed cone Engine (ACE) MBDCA (Elekta Brachy, The Netherlands) in High Accuracy level.[3] Specifically, the material composition of water was assigned to the entire phantom, so that the Total Energy Released per unit Mass (TERMA) is calculated using the mass energy absorption coefficients of water, while the density of the phantom coincided with that of RW3 using the "HU based" option of the TPS. Under these conditions, ACE interpreted the phantom material as water with the density of RW3 to accurately account for photon beam attenuation. Consequently, the fundamental dose-to-medium in medium calculation performed by ACE was equivalent to that of dose-to-water in medium, as described in AAPM TG329.[25] Similarly to the TG43-based calculation, the dose grid resolution was set to 0.5×0.5×0.5 $mm^3$.

All relevant data, including the CT image stack, treatment planning information, contoured structures, source dwell times and positions, source air kerma strength ($S_k$), as well as TG43- and ACE- calculated 3D dose distributions, were exported from the TPS in dicom-RT format for further analysis and comparison, following the workflow in Figure 1.

Treatment delivery was carried out by a remote afterloader (FlexHDR20ch, Elekta Brachy, The Netherlands). The Flexisource Ir-192 source was used with an $S_k$ of 12765.5 cGy $cm^2$/ h on the treatment delivery day. The $S_k$ of the source was based on the value on its certificate which was also verified within uncertainties (<1.5%) by the clinical user following the procedure for $S_k$ determination using a well-type chamber outlined in the IAEA TRS 492 protocol[26] and after applying the necessary decay correction. The decay-corrected treatment duration was 414.3 sec (Table 1b).

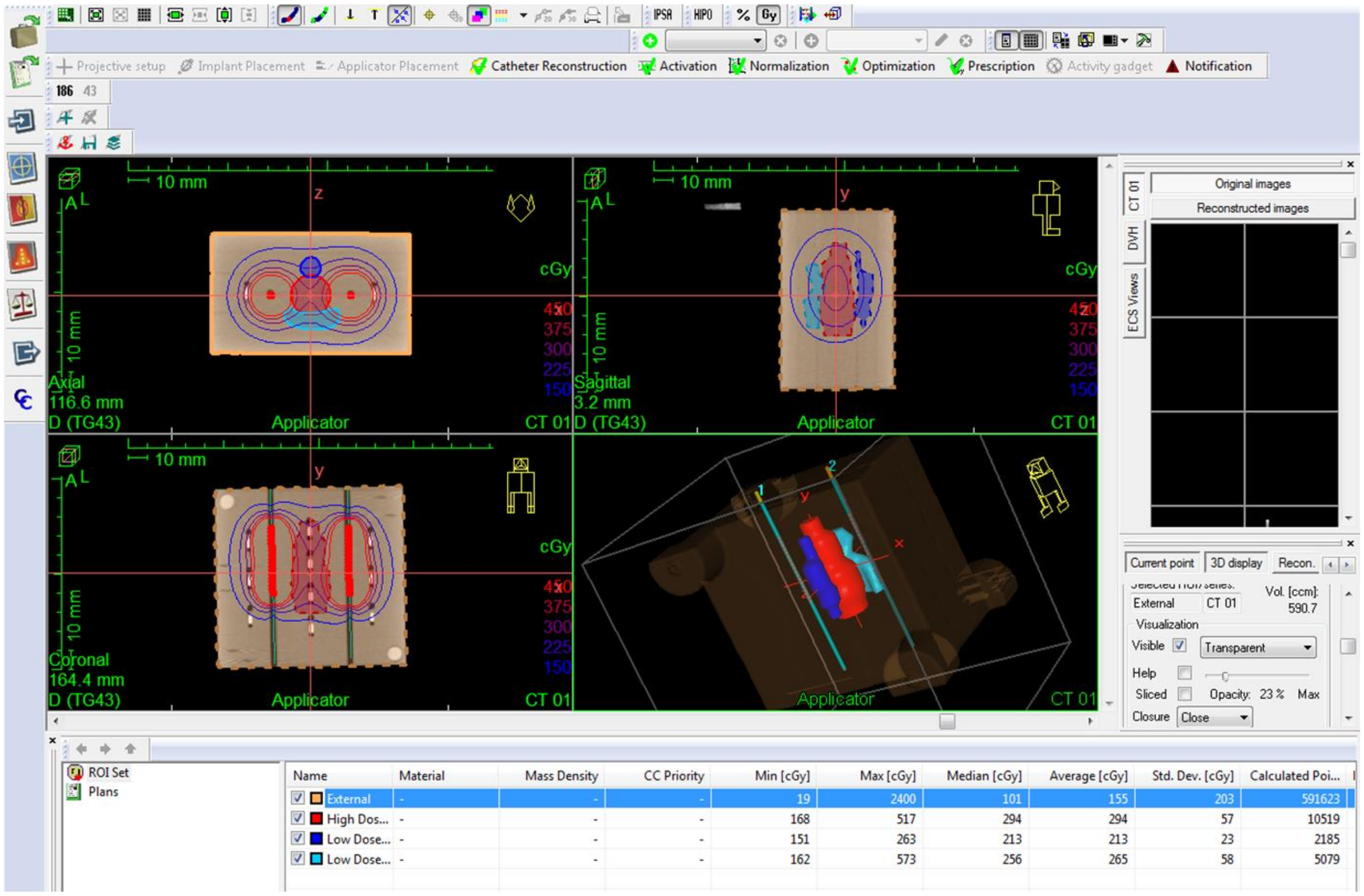

*Figure 3. Screenshot of the Oncentra Brachy v 4.6 (Elekta Brachy, The Netherlands) TPS considered in the specific implementation as a pilot study. Isodose lines (calculated using the TG43 algorithm) are superimposed on reconstructed planning CT images. A 3D representation is also shown (bottom-right panel). The plan consists of 36 source dwell*

*positions (depicted by red marks) distributed on two catheters. Three structures were contoured for plan and dose evaluation purposes; red contour: "high dose region"; blue: "low*

### 2.4 Experimental dosimetry audit test

The experimental dosimetry audit test, within the context of this pilot study, involved point dose measurements using ten OSLDs and 2D dose maps on the two film planes. Details of the corresponding dosimetry protocols are given in the following sections.

#### *2.4.1 OSL dosimetry protocol*

A batch of 100 myOSLchip™ dosimeters[27, 28] (RadPro International, Germany) and the dedicated reader and bleacher were employed for point-approximating dose measurements in the phantom described above. The active volume of this commercially available OSLD system is made of BeO (density of 2.85 g/cm$^3$) housed in a plastic case. The "High Accuracy" dosimetry protocol outlined in AAPM TG191 report[29] was followed for dose-to-water, $D_w$, in Ir-192 beam quality, $Q$, determination according to:

$$D_w = M_Q N_{D,w,Q_{ref}} k_{s,i} k_L k_Q k_\theta k_f k_{vol} \quad (1)$$

where $M_Q$ is the OSLD readout signal, $N_{D,w,Q_{ref}}$ is the calibration coefficient of the batch in beam quality $Q_{ref}$ = 6MV, $k_{s,i}$ is the individual sensitivity correction factor, while $k_L$, $k_Q$, $k_\theta$, $k_f$, and $k_{vol}$ are the linearity, beam quality, orientation, temporal signal fading and volume averaging correction factors, respectively.[29] These corrections are well-defined in the TG191 report[29] while a meticulous dose-response characterization and correction factors determination for this OSLD batch and reader along with specific implementation considerations were included in a previous publication.[30]

Briefly, nine OSLDs served as the "standards" (according to the terminology adopted in TG191) [29] for the determination of a session-specific calibration coefficient, $N_{D,w,Q_{ref}}$ (Eq (1)). For this purpose, a dose of 0.5 Gy was delivered by a reference 6MV 10×10 cm$^2$ field (VersaHD, Elekta, UK) in a modified RW3 slab phantom, under the en face orientation.[30] The reference dose was verified by a Farmer ionization chamber placed in the same slab phantom, and after applying all necessary corrections according to IAEA TRS 398.[31] Irradiation, storage and readout conditions of the standards were closely matched with the ones considered as experimentals to mitigate temporal signal fading effects and minimize all necessary correction factors for dose-to-water determination,[29, 30] i.e., $k_f$ in Eq (1) was unity. Linearity correction factors, $k_L$ (Eq (1)), were adopted from our recent publication[30] for the same batch and readout parameters. Hereinafter, the terms "standards" and "experimentals", referring to OSLDs are used in the context and terminology defined in the TG191 dosimetry protocol. [29]

The brachytherapy phantom is designed to accommodate up to ten OSLDs in sockets designed specifically for this OSLD model and size (Figures 2a,c). The detectors can be positioned in the sockets via a snug fit with the centroid of their active volume lying on the plane defined by the two catheters (interface between slabs 3 and 4, Figure 2a). OSLD orientation with respect to the catheters is always en face, as in the irradiation of the standards and thus minimizing the orientation dependence.[30] It is noted that an edge-on irradiation is not feasible in this phantom. Consequently, a $k_\theta$ of 1.000±0.014 was considered (Eq (1)) based on the results of a previous study.[30]

Regarding beam quality, $Q$, dependence of the OSL dose-response, it was accounted for by the $k_Q$ factor,[29] defined as the ratio of calibration coefficients, $N_{D,w,Q}$ in the two relevant beam qualities:

$$k_Q = \frac{N_{D,w,Ir}}{N_{D,w,6MV}} = \frac{D_{w,Ir}/M_{Ir}}{D_{w,6MV}/M_{6MV}} \tag{2}$$

where $D_{w,Ir}$ and $D_{w,6MV}$ refer to the delivered dose-to-water by the beam quality of Ir-192 (specifically at the location of the detector and the irradiation plan considered, experimental conditions) and 6MV (calibration conditions), respectively, while $M_{Ir}$ and $M_{6MV}$ represent the corresponding OSLD responses. To estimate the $N_{D,w,Ir}$ for the specific experimental implementation, a side study was designed and implemented. One OSLD was irradiated by a single dwell position of an Ir-192 HDR afterloader in air at a source-to-detector distance of 10.2 cm. The detector was placed in-between two RW3 slabs, 2 mm thick each, to achieve a charged particle equilibrium while primary photon beam attenuation is negligible. Measurements were repeated five times (using different OSLD detectors) for reproducibility considerations and the mean OSLD response was considered as the $M_{Ir}$ (Eq (2)).

Reference dose-to-water, $D_{w,Ir}$ (Eq (2)), was measured by an ionization chamber and was also verified by MC simulations. More specifically, the reference point of a Semiflex 3D (PTW, Germany) ionization chamber with a 2-mm buildup cap was placed at 10.2 cm from the source, replicating the corresponding setup for the OSLDs. A beam quality correction factor of unity (within uncertainties) was applied to the chamber's reading which was determined based on the methodology presented in Thrapsanioti et al.[32] Regarding MC-based verification of the measured $D_{w,Ir}$, the Flexisource brachytherapy source was simulated in air. The OSLD was modeled according to the blueprints provided by the vendor. Details are included in Drakopoulou et al.[30] Simulation geometry replicated the actual experimental setup. Dose-to-water was scored using $4\times10^9$ initial histories, resulting in a statistical uncertainty of <1%. MC results were converted from MeV/g to Gy using the $S_k$ and irradiation time recorded during measurements.

Similarly, irradiation of nine standards by a 6MV 10x10 $cm^2$ reference field was carried out to determine the $N_{D,w,6MV}$ in Eq (2). It is noted that an experimental approach to determine the $k_Q$ correction factor includes any potential intrinsic relative energy response.[33] This approach allowed for $k_Q$ determination for the primary Ir-192 photon beam quality only. Inevitably, beam quality varies with distance from the Ir-192 source inside the phantom due to the varying scattered photons contribution to the delivered dose.[34] However, all OSLDs are strategically placed close to the catheters (⩽2 cm, Table 1a) where primary photons are dominant. The rationale for applying a fixed $k_Q$ is discussed in more detail in section 4.

The last correction applied to the OSLD measurements of the pilot study is $k_{vol}$ (Eq (1)) to account for the volume averaging effect. In contrast to all other corrections (which can be determined once and applied in all subsequent audit tests), $k_{vol}$ is plan- and position-specific due to the different dose gradients that may be employed in a given plan and distance from the catheters. Inevitably, this correction factor must be calculated specifically for the treatment plan of this pilot study, according to the workflow outlined in Figure 1. This was performed using MC simulation with MCNP v.6.2 code.[35] However, to ensure that the experimental audit test is independent of the computational one, specific MC calculations were performed with scoring voxel sizes and locations relevant to $k_{vol}$ determination only which is a relative dose calculation procedure, i.e., disregarding the air kerma strength, and absolute source dwell times. Moreover, the geometry considered in this step is analytical, i.e., it did not rely on the CT image stack but replicates the geometry and material properties of the physical phantom and the OSLDs. It should be noted that dwell positions and time weights in MC simulation were identical to those employed for the irradiation, as defined in the corresponding dicom-RT file (Figure 1). The reason for selecting an analytical over a CT-based geometry is two-fold: (i) simulation efficiency is significantly enhanced and (ii) results are not affected by HU maps, i.e., $k_{vol}$ for OSL dosimetry is independent of the computational dosimetry CT-based dose maps. The correction factor was finally determined as the ratio of the dose scored in the centroid of each active volume (point-approximating scoring voxel dimensions of 0.2×0.2×0.2 $mm^3$) and the mean dose to the entire OSLD active volume (4.65×4.65×0.5 $mm^3$).

It is noted that the myOSLchip detectors, reader and readout protocol were calibrated and configured for clinical dosimetry in radiotherapy.[30] Thus, the sensitivity of the photomultiplier was adjusted to avoid signal saturation effects.[29] Based on the remarks of a previous study,[30] the imaging dose delivered to acquire the planning CT scan results in negligible signal (well below uncertainty levels) during OSLD readout.

### 2.4.2 Film dosimetry protocol

An EBT3 film batch (Ashland Inc., NJ, USA) was procured (Lot #10312201) and used throughout this pilot study. The dose-response curve was determined by following a rigorous calibration procedure.[36] In specific, 12 pieces, 4×4 $cm^2$ each, were cut and irradiated at a depth of 5 cm in an RW3 slab phantom at doses to water in the range of 0 – 14 Gy, using a 6MV 10×10 $cm^2$ flattened beam (VersaHD, Elekta, UK). Delivered doses were validated by a Farmer ionization chamber and considering the same setup, within the same irradiation session. All film pieces were scanned using a flatbed scanner (EPSON V850 Pro) 24 hours, 72 hours, and 7 days after irradiation to derive the corresponding calibration curves. Film handling, storing, and imaging was performed following the AAPM TG235 recommendations.[36] Imaging pixel size was 0.17×0.17 $mm^2$, i.e., a resolution of 150 dpi was selected, and all image correction routines were disabled. Each irradiated piece was positioned in the central area of the scanner and images were acquired in transmission mode. To minimize random noise in the image, each piece was imaged five times and the mean image was calculated.

The dose-response calibration curve was determined based on a multi-gaussian triple channel method[37] by employing the Radiochromic.com (Radiochromic Ltd, Spain) film analysis software and the relevant dosimetry protocol.[38] More specifically, images were imported to the software for analysis and a region of interest (ROI) was selected around the central area of the irradiated film piece. The software automatically fits a multi-gaussian model to determine the calibration curve for the given lot. Pre-irradiation film images are optional in this dosimetry protocol and, thus, not considered in order not to further burden the overall workload of the procedure (Figure 1).

The phantom can accommodate up to two dosimetry films at a distance of 1 cm from the catheters' plane (Figure 2 and Table 1a). In order to ensure film stability and facilitate registration of measurements to the CT dicom coordinate system, three metal pins per film were fixed at random locations near the film edges (Figure 1d). The pins lance through each film creating holes while their centroids can be accurately localized in the CT image stack, exploiting the increased HUs. A rigid 2D/3D transformation is established with the pin centroids (in the CT image stack) and hole centers (on the scanned film images) serving as the fiducial markers to guide the registration. Imaging dose induces negligible optical density as compared to the signal corresponding to the prescription dose and the associated uncertainties in film readout. This methodology was introduced and validated after developing all relevant image processing routines in Matlab (The MathWorks, Inc, USA) in a previous publication.[39]

Images were then converted to 2D dose distributions by applying the corresponding triple channel calibration derived in Radiochromic.com. Dose maps were exported from the software in dicom-RT format for further analysis and comparison in Matlab. EBT3 films are not considered to exhibit considerable relative energy dependence[40, 41] and thus no further processing was required to derive the dose-to-water in RW3 distribution. However, it is recommended that when the dose-response calibration is determined in 6MV photon beams and applied in Ir-192 dosimetry procedures, a low dose cut-off threshold is considered, given the dose-dependent nature of the uncertainties involved.[42] No normalization was applied; dose distributions were compared in terms of absolute dose-to-water values.

## 2.5 Computational dosimetry audit test

A computational dosimetry audit test was conducted independent of the experimental one but was relevant to the same treatment procedure and workflow of the pilot study. This test mainly focuses on the validation of the dose calculation engine and associated algorithms, incorporated in the TPS being audited. For this purpose, all treatment planning data relevant to the experimental dosimetry audit test were exported from the TPS in dicom-RT format and imported to the BrachyGuide v.2.0 software[43] for analysis. This software allows for the generation of the input file for MCNP v. 6.2 simulation (including the voxelized geometry, material compositions, source dwell positions and relative times) in order to perform dose-to-water in medium calculations, implementing a well-documented procedure for reference dose determination using this radiation transport MC code.[15, 20, 43, 44] To enhance the efficiency of MC simulation, the in-plane CT resolution was halved, while maintaining the CT slice thickness, resulting in scoring voxels of $0.54×0.54×0.6$ mm$^3$. Given the initially high resolution of the CT images, this procedure did not introduce systematic bias into the results and allowed for the determination of Dose Volume Histograms (DVHs) and comparison with the ones calculated by the TPS. A total of $4.8×10^9$ histories were used, while the source was represented by a phase space file of $8×10^7$ initially emitted photons emerging from the Ir-192 Flexisource. 3D absolute dose-to-water maps were derived using a tally multiplier proportional to the Total Reference Air Kerma (TRAK) available in the treatment plan, to convert MC output results to Gy. This branch of the audit procedure resulted in reference geometry- and plan- specific 3D dose distribution. This methodology and software involved have been repeatedly validated in previous publications and more details are given therein.[14, 15, 20, 43, 44]

Calculations were performed by a local workstation equipped with 24 cores clocked at 2.3 GHz. For this pilot study, the wall clock time needed to achieve a statistical uncertainty of up to 2.5% within a radial distance of 4 cm from the catheters was 84 hours (Table 2).

## 2.6 Uncertainty analysis

The uncertainty budget for all three dosimetry procedures is given in Table 2. Wherever possible, the uncertainty was estimated after conducting relevant tests and side studies. In several cases, the uncertainty was adopted from the literature, as the best available estimate.

OSLD accuracy relies on a set of correction factors that need to be applied to account for dose-response dependencies (section 2.4.1 and Eq (1)). A detailed characterization and a corresponding uncertainty analysis has been included in a previous publication.[30] The beam quality and the volume averaging correction factors, $k_Q$ and $k_{vol}$, respectively, are only relevant to the present study, however. They were determined by carrying out measurements and calculations (see section 2.4.1) and the corresponding uncertainty is allocated in Table 2. More specifically, uncertainty in $k_Q$ was determined as the combined uncertainty for all terms in Eq (2). One standard deviation of the repeated OSLD measurements was ascribed as uncertainty for $M_{Ir}$ and $M_{6MV}$ (1.1% and 0.8%, respectively). The overall linac output calibration uncertainty (1.5%, Table 2) was adopted as uncertainty in $D_{w,6MV}$, while uncertainty in $D_{w,Ir}$ using an ion chamber and simulations was based on a previously published study[32].

Regarding film dosimetry, the uncertainty in dose-response calibration curve is dominant (Table 2). It is dose dependent, increasing at lower dose levels. At the dose range considered in the experimental study (>1.5 Gy), uncertainty is limited to 3.5%. This level was established based on a detailed characterization performed in a previous study.[39] Relative energy dependence is not considerable in this dose range[39] (section 2.4.2) but may be significant at the lower dose areas,[42] if the corresponding uncertainty allocated in the budget is not adequately large.

Regarding the computational dosimetry test, dosimetric uncertainties given in Table 2 involve statistical (type A) and systematic (type B) components. The latter was adopted from AAPM TG138 repoort.[45] It is noted that uncertainty in source strength is not relevant since the TPS algorithms and MC methods both consider the nominal (decay-corrected) $S_k$ to perform dose calculations. A spatial uncertainty component is not applicable since the coordinate system and the scoring grid are identical across all dose distributions. Inaccuracy in catheter digitization and reconstruction cannot impact the results of this computational dosimetry test (Table 2).

*Table 2. Uncertainty analysis for the experimental (OSLDs and film measurements) and computational dosimetry (MC calculations) audit tests, according to the procedure described in section 2 and Figure 1.*

| Source of uncertainty | Type | Dosimetric uncertainty (%) | Spatial uncertainty (mm) | Comment/ reference |
|---|---|---|---|---|
| **OSL dosimetry** | | | | |
| Source air kerma strength | B | 1.5 | - | [45] |
| Dose-response calibration (linac output) | B | 1.5 | - | Overall linac output calibration uncertainty |
| Session specific $N_{D,w}$ | A | 1.4 | - | Estimated as one standard deviation of the mean response of the nine standards |
| Reader stability | A | 0.8 | - | Reference light variation of the OSLD reader.[30] |
| Element sensitivity factor, $k_{s,i}$ | B | 0.7 | - | [30] |
| Linearity correction factor, $k_L$ | B | 1.7 | - | [30] |
| Angular correction factor, $k_\theta$ | B | 1.4 | - | [30] |
| Volume averaging correction factor, $k_{vol}$ | A | 0.8 | - | Combined statistical uncertainty of the performed simulations in an analytical geometry |
| Beam quality correction factor, $k_Q$ | B | 2.5 | - | Determined by conducting a side study. It is the combined uncertainty for all terms in Eq (2).[30, 32] |
| Catheter reconstruction | B | - | 0.6 | One CT image slice thickness |
| Active volume localization | B | - | 0.5 | [30] |
| Centroid determination | B | - | 0.12 | [30] |
| Total standard uncertainty (k=1) | A+B | 4.4 | 0.8 | |
| Total expanded uncertainty (k=2) | A+B | 8.8 | 1.6 | |
| **Film dosimetry** | | | | |
| Source air kerma strength | B | 1.5 | - | [45] |
| Dose-response calibration | B | 3.5 | - | Dose dependent uncertainty, which increases with decreasing dose.[39] |
| OD readout reproducibility | A | 0.3 | - | [39] |
| Scanner reproducibility | A | 0.25 | | [46, 47] |
| Scanner homogeneity | B | 0.2 | - | [46, 47] |
| Catheter reconstruction | B | - | 0.6 | One CT image slice thickness |
| Registration | B | - | 1.5 | [39] |
| Total standard uncertainty (k=1) | A+B | 3.8 | 1.6 | |
| Total expanded uncertainty (k=2) | A+B | 7.6 | 3.2 | |
| **Computational dosimetry** | | | | |
| MC statistics | A | <2.5 | - | Indicative statistical uncertainty for a voxel located 4 cm away from the catheters |
| Ir-192 source emission, transport code, interactions and scoring cross sections | B | 0.3 | - | Indicative quadrature combination of systematic uncertainties for a voxel at 5 cm from the source.[45] |
| Spatial co-registration | B | - | - | Not applicable as MC distributions are directly determined in the CT coordinate system |
| Catheter reconstruction | B | - | - | Not applicable as catheter reconstruction spatial errors cannot affect computational dosimetry audit results |
| Total standard uncertainty (k=1) | A+B | <2.5 | - | |
| Total expanded uncertainty (k=2) | A+B | <5.0 | - | |

## 2.7 Dose comparisons

All measurements were compared against TPS-calculated dose distributions (TG43 and ACE dose predictions) by implementing the Local and Global 3D Gamma Index (GI) tests.[46] In particular, measurements always served as the reference dataset against which the TPS calculations were evaluated in 3D. No dose normalization was applied to any of the datasets. The 5%/1mm passing criteria were selected for the OSLDs and 4%/2mm for the films (applied both locally and globally) based on the uncertainty analysis for each system, presented in Table 2. A low dose cut-off threshold of 1.5 Gy was applied to exclude film areas of low clinical importance which are also found near the edges of the phantom. This threshold is also justified by the optimal dose-response range in Ir-192, if films are calibrated in high energy photon beams.[42]

Prior to any GI calculation, the evaluated dataset was interpolated to a finer isotropic spatial resolution of 0.1 mm, which corresponds to the distance-to-agreement (DTA) divided by a factor of at least 10, as recommended in Hussein et al.[47]

Regarding the computational dosimetry test, MC results were considered as the reference 3D dose distribution with TPS calculations (TG43 and ACE) regarded as the evaluated ones. Given that MC distribution is directly registered to the CT coordinate system and catheter reconstruction inaccuracies are not relevant, there is no spatial uncertainty component in this procedure (Table 2) and, thus, the GI test was not considered. Computational dosimetry audit test results relied on 3D local dose difference maps after applying linear interpolation to calculate the dose at the same grid of points. A local dose difference is defined as $\frac{D_{eval}(x,y,z)-D_{ref}(x,y,z)}{D_{ref}(x,y,z)} \times 100\%$, where $D_{ref}(x,y,z)$ and $D_{eval}(x,y,z)$ are the reference and evaluated doses at $(x,y,z)$, respectively.[17] Moreover, the comparison with TPS calculations was also performed in terms of DVH analyses for the hypothetical high and low dose structures created during treatment planning (Figure 3). Although the treatment plan considered was not clinically realistic, this pilot study demonstrates the usefulness and importance of incorporating 3D dose distributions in an audit test. This comparison was also carried out in BrachyGuide[43] (see section 2.5).

# 3. RESULTS

Upon completion of phantom construction, calibration of the dosimeters and study design, the developed integrated dosimetry protocol was implemented in a pilot study in order to reveal potential limitations in real world clinical settings, in the context of a remote audit test. The integrated workflow involved several procedures which were completed within ten days from the arrival of the phantom at the clinical site. The most time-consuming steps are the MC calculations (performed only after exporting the relevant dicom-RT data from the TPS, Figure 1) and the necessary time between OSLD irradiation and readout that should elapse to mitigate potential temporal signal fading effects.[30] Results of this pilot study are given in the following sections.

## 3.1 Experimental dosimetry

### *3.1.1 OSL dosimetry*

Beam quality dependence of the OSLD response was found of the order of 6% for the measurements performed in air, i.e., Ir-192 spectrum in absence of scattered photons. Given the close proximity of all OSLDs to the catheters (Table 1a), the dose component by the primary photons is dominant. Consequently, a $k_Q$ (Eq (1) and (2)) of 1.06±2.5% (Table 2) was applied to the OSLD measurements in the pilot study. The rationale for applying a fixed $k_Q$ is further discussed and justified in section 4.

The plan- and position- specific $k_{vol}$ (Eq (1)) was determined via MC-based relative dose calculations in an analytical geometry, simulating the actual voxel-based one. The corrections needed ranged between 0.985 and 1.058. The statistical uncertainty in $k_{vol}$ reached 0.8% (Table 2).

Results of the 3D Local and Global GI tests are given in Table 3. ACE and TG43 perform equally well. The only OSLD failing the Local GI test was found in a low dose region (measured dose: 1.59 Gy). Local dose differences are also given in Table 3. The mean and median dose differences are not significant (well within experimental uncertainties), indicating negligible systematic errors in the procedure.

*Table 3. Gamma Index (GI) passing rates and local dose differences relevant to the OSLDs measurements performed in the phantom as part of the implementation study. No low-dose cut off threshold applied.*

| **Evaluated distribution** | **3D GI passing rates (5%/1mm)** | | **Local dose differences (%)** | | | |
|---|---|---|---|---|---|---|
| | Local GI | Global GI | Mean | std[a] | Median | Max |
| TG43 | 9 out of 10 | 10 out of 10 | 0.1 | 5.4 | 1.3 | -9.9 |
| ACE | 9 out of 10 | 10 out of 10 | -0.4 | 6.2 | 1.3 | -11.6 |

[a]standard deviation

### *3.1.2 Film dosimetry*

Holes on the film images and corresponding metal pins in the CT image stack are well-defined with adequate contrast, allowing for an accurate localization of their centroids. The total spatial uncertainty (including catheter reconstruction) reached 1.6 mm (Table 2).

An overview of the 3D Local GI maps is given in Figure 4, for both films. A low dose cut-off threshold of 1.5 Gy was considered to exclude areas of the film close to the phantom edges and mitigate increased calibration uncertainty. Local and Global GI passing rates exceed 95% for both algorithms and both films (Table 4). Although not shown here, systematic discrepancies between measurements and calculations were noticed at the low dose areas between 1 and 1.5 Gy. This can be attributed to increased uncertainty at low doses if films are calibrated at 6 MV beams and irradiated by Ir-192,[42] and partly to the missing scatter conditions close to the edges of the phantom (relevant to the TG43 algorithm only, e.g., Figure 4b).

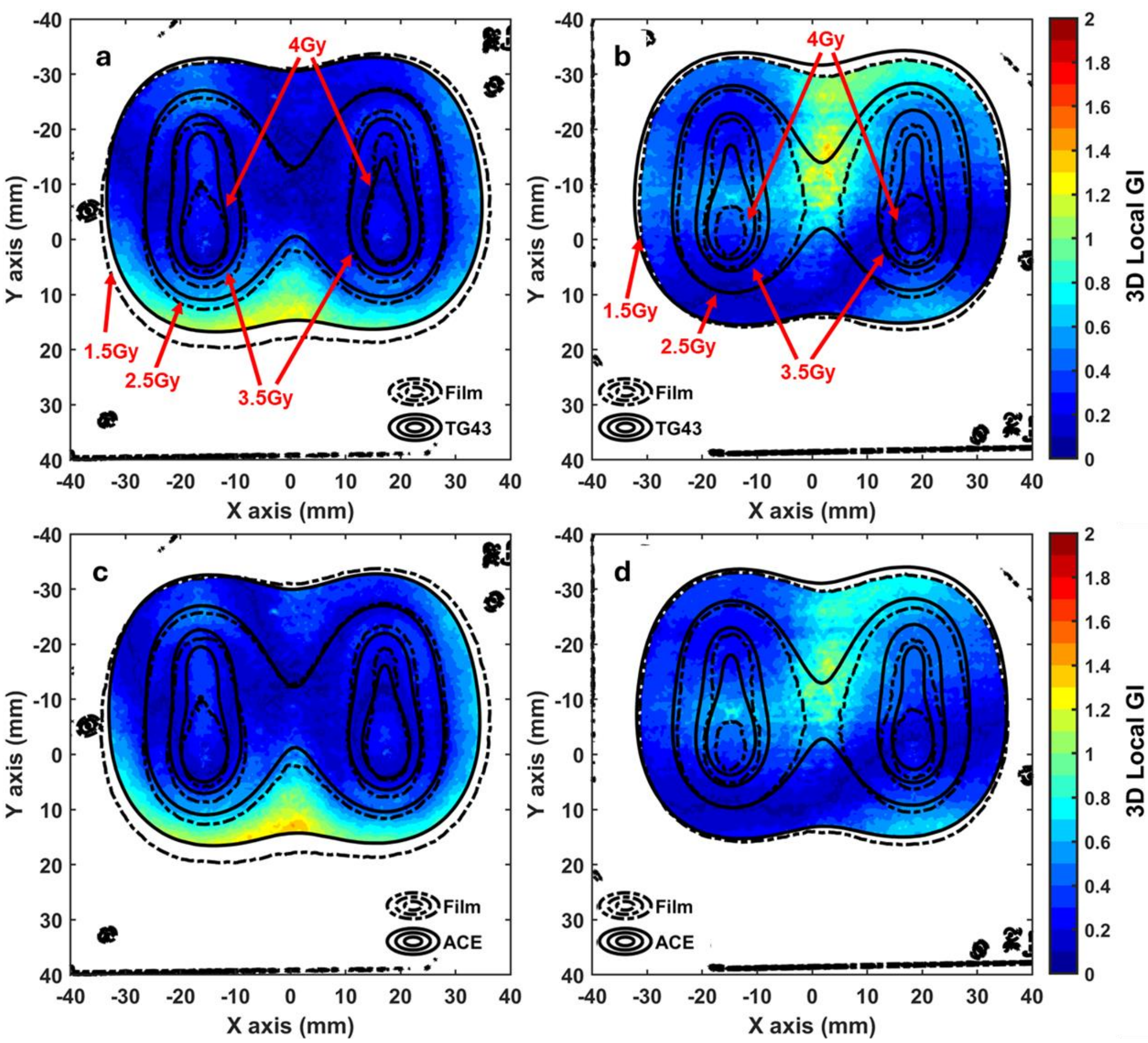

*Figure 4. Isodose lines of the measured (dashed lines) and calculated (solid lines) dose distributions superimposed on the 3D Local GI maps (passing criteria 4%/2mm). (a, b) TG43 calculations compared with the (a) posterior and (b) anterior film measurements. (c, d) ACE calculations compared with the (c) posterior and (d) anterior film measurements. The isodose levels are given in Gy in red fonts. Black marks are artifacts related to holes (for registration) or markers (for identification and orientation) present on the film images.*

*Table 4. Passing rates for the applied 3D Gamma Index (GI) tests relevant to the film measurements. Passing criteria considered were 4%/2mm with a low dose cut-off threshold of 1.5 Gy.*

| **Evaluated distribution** | **Posterior film** | | **Anterior film** | |
|---|---|---|---|---|
| | Local GI | Global GI | Local GI | Global GI |
| TG43 | 97.1% | 98.0% | 96.3% | 99.3% |
| ACE | 96.4% | 97.2% | 99.6% | 99.9% |

## 3.2 Computational dosimetry

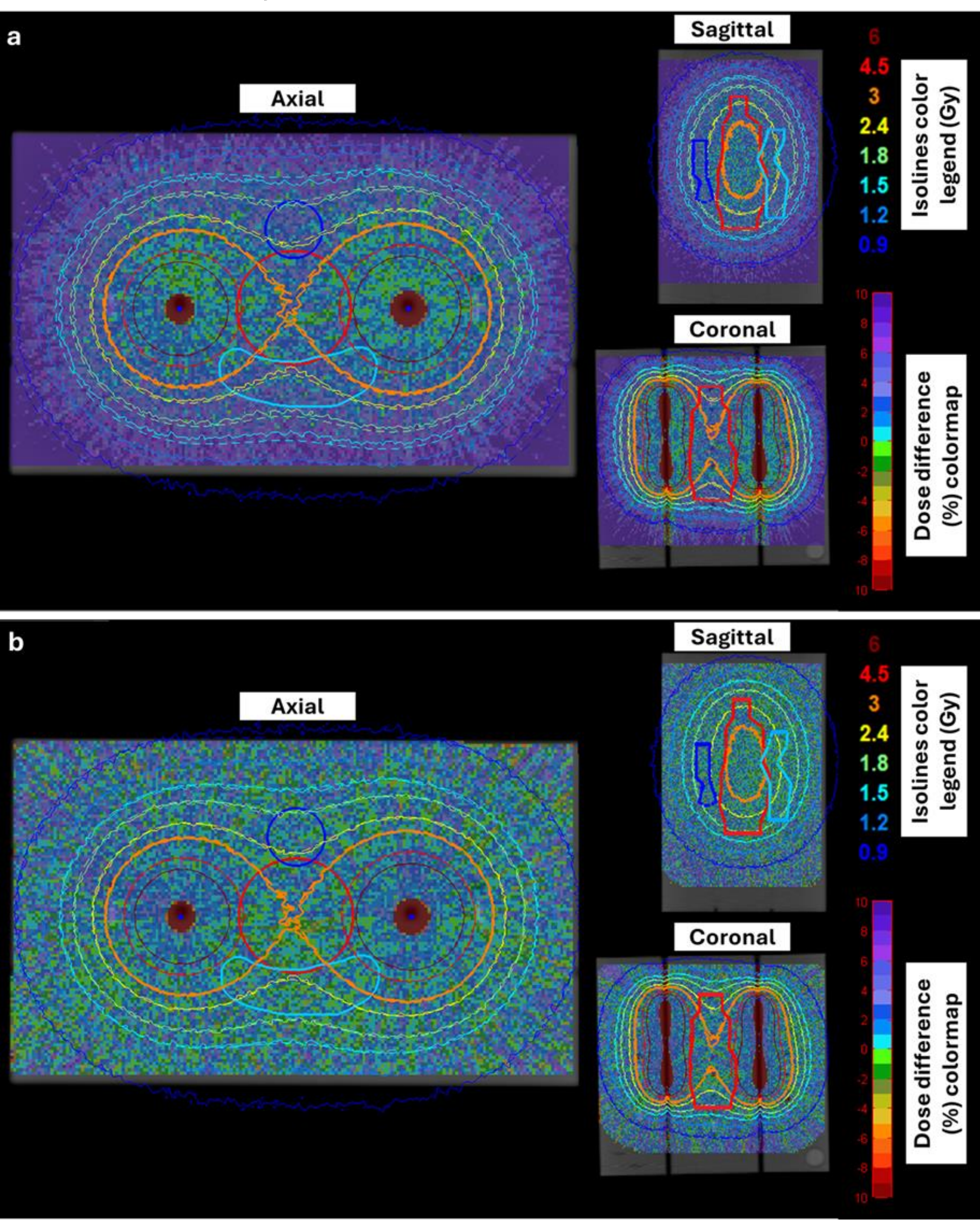

*Figure 5. Axial, sagittal, and coronal slices of the reconstructed CT image stack of the phantom. Local dose difference maps comparing MC results with (a) TG43 and (b) ACE predictions are superimposed. The corresponding isodose lines for MC (solid lines) and TPS*

*(dashed lines) distributions are also shown. Color legend for the structures: red: high dose structure, blue: low dose structure 1; turquoise: low dose structure 2, as defined in Figure 3.*

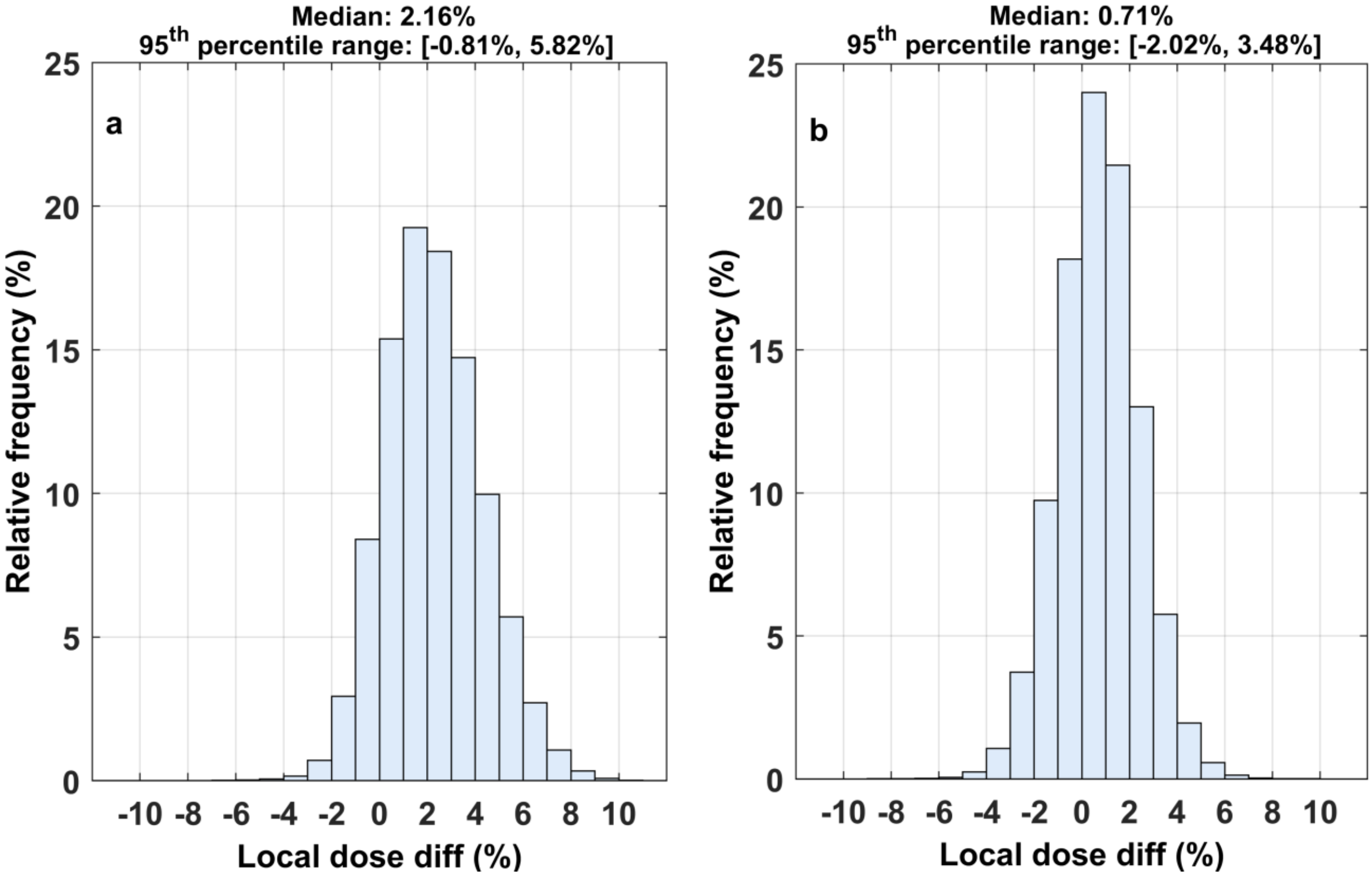

*Figure 6. Local dose difference histograms (1% wide bins) for the entire 3D phantom volume to compare MC dosimetry results with (a) TG43 and (b) ACE dose predictions in the pilot study. High and low dose cut-off thresholds of 24 Gy and 1.5 Gy, respectively, have been applied. Median values and the 95$^{th}$ percentile range are also given.*

The computational dosimetry audit test results are summarized in Figures 5-7. Figure 5 presents three indicative slices of the geometry considered along with corresponding isodose lines and local dose difference maps. The isodose lines and dose differences between TG43 and MC dose distributions in Figure 5a demonstrate close agreement up to approximately 2 cm from the source dwell positions. Beyond this distance however, the agreement deteriorates with TG43 exhibiting a dose overestimation (of the order of 2%) that exceeds MC Type A uncertainty for the vast majority of voxels. This degree of overestimation is attributed to the disregarded effect of the RW3 density by TG43, combined with the missing scatter conditions which are more pronounced closer to the edges of the phantom. Corresponding comparisons between ACE and reference MC results in Figure 5b show a general agreement within MC Type A uncertainty throughout the entire geometry, except for distinct increased differences that manifest as rays, due to the inherent ACE discretization artifacts.[15, 34] For both TG43 and ACE, the increased negative differences relative to MC results evident around the source dwell positions are artifacts associated with the dose threshold of 800% (i.e., 24 Gy in this pilot study) applied automatically by the TPS.

These remarks are confirmed in the entire 3D phantom volume, as shown in Figure 6, in which relative frequency histograms of the detected local dose differences are given in bins of 1% width. High and low dose cut-off thresholds of 24 Gy (800%) and 1.5 Gy, respectively, have been applied to exclude voxels in the vicinity of the catheters (automatically disregarded by the TPS) and those at large distances from the implant. According to Figure 6, both TG43- and ACE-based dose difference distributions appear normally distributed. Despite TG43 yielding higher dose values than MC in Figure 6a, the corresponding median dose difference was 2.16% ($95^{th}$ percentile range: [-0.81%, 5.82%]). An excellent agreement can be observed in Figure 6b between ACE and MC, with the dose difference distribution centered around zero, resulting in a median value of 0.71% ($95^{th}$ percentile range: [-2.02%, 3.48%]).

Further leveraging the 3D nature of the computational dosimetry test, the calculated DVHs for the three structures considered are presented and compared in Figure 7. Consistent with the voxel-based analysis, DVH comparison shows a good agreement between ACE and MC, whereas a notable -yet expected- shift towards higher values is observed for TG43 predictions with respect to MC for all structures considered.

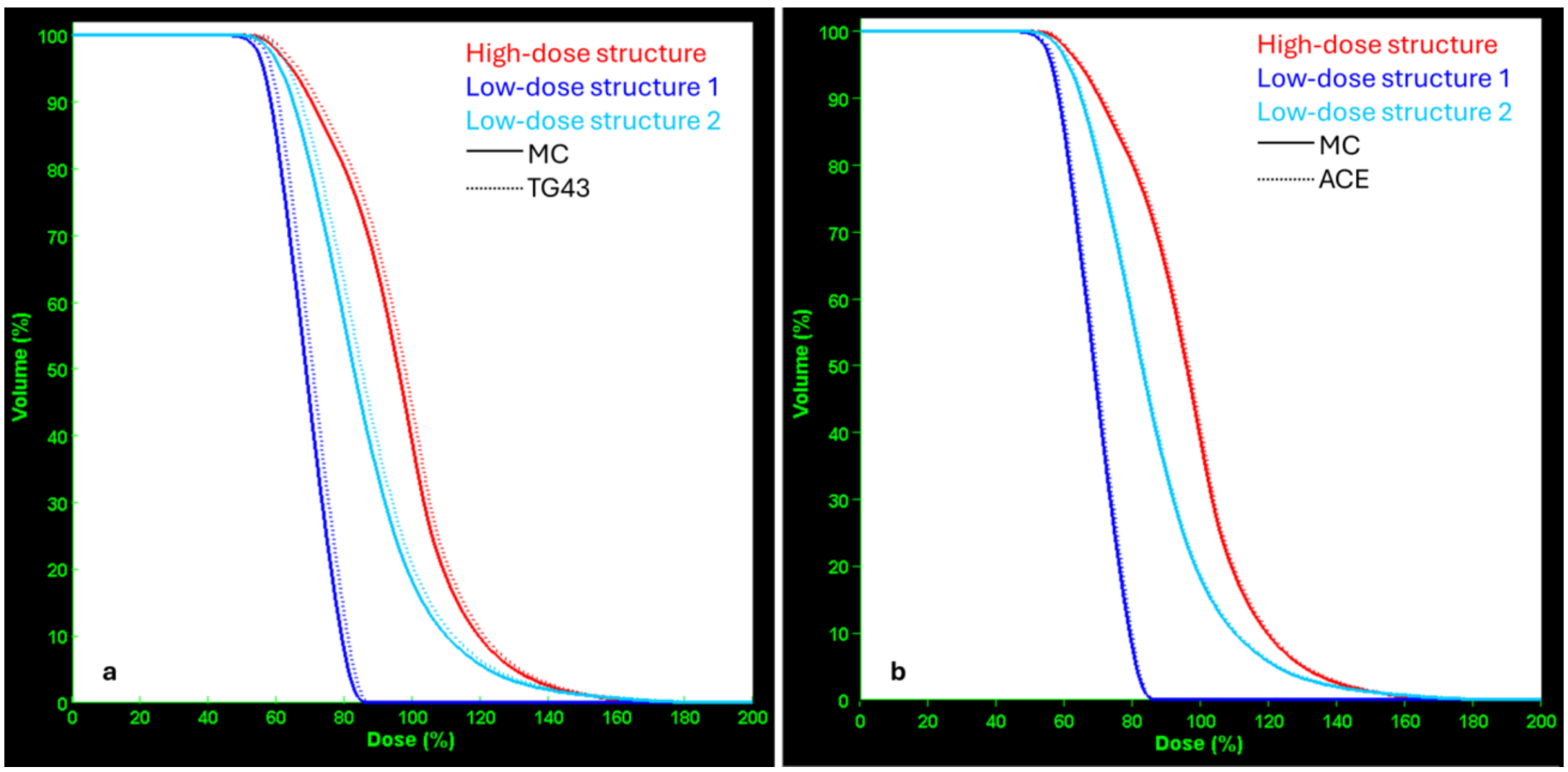

*Figure 7. Dose Volume Histograms (DVHs) calculated using MC methods compared against (a) TG43 and (b) ACE predictions for the three structures considered during treatment planning in the pilot study (Figure 3).*

## 4. DISCUSSION

An integrated workflow combining experimental and computational dosimetry was developed and implemented. Emphasis was put on the determination of a detailed uncertainty budget in order to conclude on appropriate passing criteria. Feasibility was demonstrated by implementing and completing the procedure within 10 days from phantom arrival at the site. Moreover, it was shown that both TG43 and MBDCAs can be audited using the same workflow, methodology and analysis. However, the lack of inhomogeneities in the current version of the phantom sets limits with respect to performance evaluation of MBDCAs.[34] The presented workflow cannot be regarded as a Level-2 commissioning test.[2] To the best of our knowledge, this is the first study to present an integrated workflow, combining experimental and computational dosimetry for auditing purposes.

The remote audit tests developed by IAEA[12] and RPC[13] also rely on compact phantoms made of acrylic and polystyrene, respectively. One or two plastic catheters are fixed and point dose measurements are compared against TPS predictions for a predefined simple plan designed to deliver homogeneous dose to the detector(s). Thus, simple -yet rigorous- audit procedures are available to clinical brachytherapy users. The present study builds on the same concept but includes ten point-approximating detectors, two planar dose measurements, and 3D dose calculations for a user-selected and optimized treatment plan with dose modulation and steep dose gradients, essentially constituting an end-to-end test, involving all steps of the treatment chain. Inevitably, implementation is far more complex for the auditor, but no extra effort and time are required from the user's end. On the other hand, experimental uncertainties are also larger (Table 2). In that respect, simpler dosimetry tests may be more suitable for routine source strength verification procedures, while the presented methodology could be implemented as an end-to-end benchmarking test if new equipment is installed or updated.

The main idea to integrate two fundamentally different dosimetry procedures (measurements and simulations) allows independent tests to be conducted concurrently but more importantly builds confidence in the outcome. In order to check for consistency between the two fundamentally different dosimetry tests, dose measurements (experimental dosimetry employing OSLDs and films) and MC results (computational dosimetry using MCNP) were compared using the same analysis method and GI passing criteria for comparison which served as a cross validation study. An outstanding agreement was observed, with passing rates reaching 100%. This remark implies that the developed methodology is valid, the reported uncertainties and passing criteria considered are reasonable and the equipment performs within acceptable tolerances.

To demonstrate the importance of combining independent dosimetry tests, equipment and systems, implementation of the developed procedure was repeated after deliberately introducing a systematic error in the catheter reconstruction during treatment planning. More specifically, the tip of the catheter was identified with an offset of 2 mm applied in a direction parallel to the catheters. Consequently, all source dwell positions were shifted by 2 mm. This spatial error was detected by OSLDs and films independently, yielding GI passing rates up to 60% and 83%, respectively. On the other hand, computational dosimetry results were not affected by the intentional error, as expected. The fact that OSLDs and films both resulted in unacceptably low passing rates strengthens the credibility of the experimental audit test, also providing evidence that the equipment and detectors performed as expected. In other words, a cross check between different experimental dosimetry systems and between independent tests is reassuring with respect to detecting errors, while validating the outcome and narrowing down the list of potential sources of error.

Although initially developed for personal dosimetry, OSLDs have been proven reliable in clinical radiotherapy dosimetry,[29] with applications ranging from in vivo dosimetry and MR-linac quality control checks to stereotactic radiotherapy end-to-end tests and proton beams.[27–30, 48, 49] Nevertheless, limitations with respect to dose-response have been identified which can be accounted for by increasing the relevant uncertainties. More specifically, orientation and beam quality dependences along with non-linearity are dominant sources of uncertainty allocated in the uncertainty budget (Table 2). Regarding $k_Q$, a value of 1.06±2.5% was determined for the primary photon beam quality. Determining a correction factor for the primary photon beam is justified by the proximity of all OSLDs to the catheters (Table 1a). Dose rates by primary and scattered photons are equal at 6 cm from a source centered in a 30-cm diameter water phantom.[34] At distances ≤2 cm (as for the OSLDs), scattered photons contribute to the total dose rate by <25% within an adequately large phantom ensuring full scatter conditions.[34] The much smaller size of the present phantom (Table 1a) further diminishes the scattered photons contribution to the total dose delivered to the OSLDs in the pilot study. Therefore, it is reasonable to correct for the dominant beam quality and apply an uncertainty large enough to account for this approximation. Jahn et al. have demonstrated that BeO-based OSLDs under-respond at kilovoltage photon energies.[33] Thus, it is expected that the $k_Q$ value determined herein is not applicable at larger distances from the catheters or larger phantom sizes. Employing MC methods for a position-specific $k_Q$ determination will not take into account intrinsic relative energy effects related to signal generation and detection and thus simulation-based results may be biased.[33] Nevertheless, a more rigorous calibration procedure using Ir-192 beam as a reference source for the calibration coefficient determination will result in reduced uncertainties in $k_Q$.[29] Uncertainty in $k_L$ may be reduced by considering lower doses during treatment planning[30] but this will also affect delivered dose to the film, hence striking a balance is not trivial.

Film dosimetry is commonly used in remote dosimetry applications and audit tests in radiotherapy, as a well-established method and mature technology.[36] However, practical issues in implementing film dosimetry should be acknowledged. More specifically, the labor-intensive procedures for calibration and measurements, combined with relatively short-term expiration dates, are the most significant limitations.[36, 50] Spatial registration of film dose maps to the coordinate system of the TPS calculations (a critical step to allow comparison) is often challenging to implement due to the fact that fiducials guiding the registration are phantom-specific.[39] Consequently, film analysis software may not incorporate a rigorous registration step based on fiducials.[38] This is the reason for developing in-house software routines to carry out this task. Regarding Ir-192 brachytherapy applications, another limitation arises; the beam quality dependence becomes increasingly significant with decreasing dose levels.[42] This is an important parameter to consider when designing a phantom, deciding on the prescription dose or evaluating the results. As with OSLDs, the effect can be mitigated if the calibration protocol involves calibration in a reference Ir-192 beam, which however may come at the expense of increased uncertainty in $N_{D,wQ_{ref}}$.[29]

The computational dosimetry methodology implemented in this study is not novel. It has been developed mainly for commissioning purposes, especially for MBDCAs.[14, 15, 17, 43] Reference dose distribution datasets can be created for digital phantoms to test the relevant accuracy of contemporary dose calculation engines. This work, however, has demonstrated the feasibility of conducting a computational dosimetry procedure for any plan and TPS calculation originally created for a physical phantom in the context of an experimental dosimetry audit. MC calculations allow for rigorous validation of the TPS-calculated dose distribution in 3D but disregard other steps of the treatment chain e.g., potential errors in source air kerma strength, catheter digitization and reconstruction, file transfer and communication, source spatial step and timer, etc. Thus, the role of computational dosimetry is complementary. Despite the fact that no variance reduction techniques were implemented, simulation efficiency was adequate to obtain a 3D dose distribution in a voxelized geometry and an acceptable uncertainty within 3.5 days using a standard workstation.

Validation of the TG43 and ACE algorithms available in the Oncentra Brachy TPS was out-of-scope for this study. Both algorithms have been validated in previous works employing the same TPS.[21, 39, 43] In the present study, comparison was performed only to demonstrate feasibility of the developed integrated workflow. Implementing a highly complex procedure (Figure 1) in a validated clinical system and workflow, serving as a pilot study, increases the validity and credibility of the presented methods and analyses. The detected systematic dose over-estimation of TG43 calculations, as compared to MC results (Figures 5a, 6a and 7a), suggests that the missing scatter conditions and phantom density induce a systematic discrepancy of 2.16% (median value within the 1.5-Gy isodose, Figure 6a) which should be considered when evaluating the results of a TG43-based dose calculation engine. Alternatively, this correction can be included in the workflow of the audit test, applicable only for TG43 algorithms.

Despite the high passing rates achieved in the pilot study, several limitations related to applicability and generalizability should be considered. The phantom does not offer the option for accommodating material inhomogeneities, such as air cavities or bone equivalent pieces.[39] This is an important omission for audit tests developed specifically for MBDCAs. However, incorporating small volumes of inhomogeneities in the current phantom design does not require extensive modification and could be considered as future work. Another noteworthy limitation is the incompatibility of the phantom with brachytherapy applicators. The reason for targeting interstitial brachytherapy only (i.e., a phantom compatible with plastic catheters) was based on the divergent designs and sizes of the applicators that are currently available in clinical practice. In other words, one phantom design that fits all is not applicable. Designing and validating applicator-specific phantoms was out-of-scope for this work. Furthermore, the phantom was designed to accommodate up to two catheters. Clinical applications, however, typically involve 6-20 catheters depending on the treatment site. In that respect, the developed procedure is not suitable for auditing clinical plans. This was decided on the grounds of developing a phantom compact in size, suitable for postal dosimetry, as well as keeping the required time for treatment planning and delivery, and MC calculations to a practical minimum. It is noted that other audit protocols rely on a limited number of catheters in a compact phantom as well.[9, 12, 13] Experimental uncertainties (Table 2) can be considered acceptable but there is room for further optimization of the developed protocol. In particular, this audit test would benefit from introducing a calibration procedure at a reference Ir-192 beam, especially in a secondary standard laboratory. This would guarantee a significant reduction in the combined overall uncertainty and thus result in a more rigorous audit procedure. Lastly, control detectors to monitor transport dose or potential unintentional irradiation were not included in this pilot implementation (due to the geographical proximity between the institutions involved) but, as good practice, will be considered in future audits.

The current pilot study was focused on Ir-192 HDR brachytherapy. With minor adjustments, however, it can be implemented in Co-60 brachytherapy modalities, provided that the beam quality dependence of the dosimeters is well-established and the phase space of the corresponding clinical source is available or can be generated for enhanced simulation efficiency. Electronic brachytherapy is being increasingly adopted in clinical practice.[16] It involves, however, low kilovoltage photon beams for which the RW3 cannot be considered water equivalent.[24] Moreover, OSLDs in such beam qualities demonstrate potentially unacceptable energy dependence.[33] For these reasons, it is expected that the developed protocol is not suitable for audit tests in electronic brachytherapy.

## 5. CONCLUSIONS

An integrated remote dosimetry audit test for Ir-192 interstitial HDR brachytherapy was developed and implemented in a pilot study. The novelty of this procedure mainly lies in the combination of experimental and computational dosimetry testing, independently, in a well-established unified protocol for the same user-selected treatment plan. The workload for the clinical user is minimal as only imaging, planning and delivery are performed on-site.

At the core of this procedure lies a phantom, compact in size, which can accommodate up to ten OSLDs and two films. Computational dosimetry is performed using BrachyGuide and the MCNP MC code package.

The uncertainty analysis resulted in acceptable uncertainty levels for all three dosimetry procedures involved. Wherever necessary, appropriate correction factors were determined. The passing criteria considered were tailored to the combined uncertainty for each system/procedure.

Point-approximating and 2D dose measurements, as well as MC-based 3D dose distributions were compared against TPS calculations (both TG43 and an MBCA). The audit procedure allows for Local and Global GI tests to be performed, while local dose difference maps and DVHs are evaluated in the 3D volume of interest. An intentional spatial error of 2 mm in catheter tip identification, applied parallel to the catheters, was detected by both OSLDs and films while being irrelevant in a computational audit test.

Despite the labor-intensive workflow for the auditing institution, the developed protocol is suitable for remote Ir-192 audit tests. Overall results of this work highlight the advantages of a multi-dosimetry approach for comprehensive and rigorous auditing programs.

**Acknowledgments**

Elekta Brachytherapy (Veenendaal, The Netherlands) is gratefully acknowledged for providing Oncentra Brachy v.4.6, for research purposes.

Theocharis Pappas is acknowledged for his technical assistance in phantom design and construction.

**Conflict of Interest Statement**

The authors have no relevant conflicts of interest to disclose.